\documentstyle[epsf]{mn}

\title[Evolution of Galaxy Clustering]
      {Evolution of Galaxy Clustering}
\author[J.S.Bagla]
       {J.S.Bagla\\
        Institute of Astronomy, University of Cambridge, Madingley Road,
        Cambridge CB3 0HA, U.K.\\
        E-mail: jasjeet@ast.cam.ac.uk}

\pagerange{\pageref{firstpage}--\pageref{lastpage}}
\pubyear{1998}

\begin{document}
\label{firstpage}

\maketitle

\begin{abstract}
We study the evolution of correlation function of dark matter halos in
the CDM class of models.  We show that the halo correlation function
does not evolve in proportion with the correlation function of the
underlying mass distribution.  Earliest halos to collapse, which
correspond to rare peaks in the density field, cluster very strongly.
The amplitude of halo correlation function {\it decreases} from its
initial, large, value.  This decrease continues till the average peaks have
collapsed, after which, the amplitude grows at a slow rate.  This
behaviour is shown to be generic and the epoch of minimum amplitude
depends only on the {\it rms}\/ fluctuations in mass at the relevant
scale and, to a much smaller extent, on the slope of the power
spectrum at that scale.  We discuss the relevance of this result for
interpretation of observations of galaxy and quasar clustering.
\end{abstract}

\begin{keywords}
Galaxies : Formation -- Cosmology : Theory -- Early Universe,
Large Scale Structure of the Universe
\end{keywords}

\section{Introduction}

It is believed that structures like galaxies and clusters of galaxies 
formed by gravitational amplification of small perturbations.  This
implies that clustering in the mass distribution in the Universe
always increases with time.  In this model, galaxies form in highly
over-dense halos of dark matter.  Evolution of galaxy clustering and
its relation to the clustering in the underlying mass distribution is an
important question that needs to be addressed before we can correctly
interpret the observations of galaxy clustering.  In this
paper we discuss one approach to this problem and comment on the
relevance of our results for interpretation of observations of galaxy
clustering and its evolution. 

Clustering properties of halos have been studied by many authors (See,
for example, Gelb and Bertschinger~(1994b)).  These studies show that
there is no simple, scale independent relation between the correlation
functions for the mass distribution and halos.  Evolution of
clustering properties of halos has been studied by Brainerd and
Villumsen~(1994).  They computed the halo-halo correlation function
and found that it evolves very slowly in most cases.  In some cases
they found a {\it decreasing}\/ phase before the amplitude of the halo
correlation function starts growing.  Recently, some authors have
studied the effects of a decreasing correlation function, and an epoch
of minimum amplitude of clustering, on the angular correlation
function of galaxies \cite{angcorr}.

In this paper we argue that the halo correlation function, for halos
above a given mass, has a generic behaviour:~its amplitude is very
large at early times, and decreases very rapidly.  The amplitude
reaches a minima and then increases at a slow rate. 

The paper is organised as follows: \S{2} outlines a model for
evolution of halo clustering.  In \S{3} we describe the numerical
simulations used for testing the model.  This section also contains
details of our approach for identifying halos and calculation of the
halo correlation function.  Results are presented in \S{4} and in
\S{5} we discuss the generalisation to galaxy clustering and the
relevance of our results for interpretation of observations.  We
summarise the main conclusions in \S{6}.

\section{The Halo Grail}

In most models of structure formation, the initial density field is
assumed to be a Gaussian random field.  Gravitational instability
leads to amplification of density perturbations, and over-dense regions
collapse to form virialised halos.  If the threshold density contrast
for formation of halos is much larger than the {\it rms}\/ dispersion in
density contrast, as is the case at early times, only a few rare peaks
collapse into halos.  These peaks are expected to cluster strongly
\cite{bbks} compared to the almost smooth mass distribution, and
are poor tracers of the underlying mass distribution.
On the other hand, if the threshold is less than, or comparable to,
the {\it rms}\/ dispersion, almost all peaks collapse to form halos
and the halo number density traces the mass distribution at scales
(much) larger than the typical inter-halo separation.  Thus we expect
the halos to become ``better'' tracers of the mass distribution with
time.  In the following discussion, we quantify these statements and
present a simple model for evolution of the halo correlation function.

Consider the distribution of halos of mass $M$, or larger, before
typical halos of this mass have collapsed.  To quantify this, we first
define a {\it bias}\/ parameter $\nu(M,z) = \delta_c/\sigma(M,z)$, where
$\sigma(M)$ is the (linearly extrapolated) {\it rms} dispersion in the
density contrast at mass scale $M$ and $\delta_c$ is the linearly
extrapolated density contrast at which halos are expected to collapse
and virialise.  We use the value $\delta_c = 1.68$, obtained from the
spherical top hat collapse model \cite{gunngott}.  Most halos of mass
$M$ collapse when $\nu(M) \approx 1$.  At early times, $\nu \gg 1$, and
the number density of collapsed halos is very small (compared to, say,
$\bar\varrho/M$, where $\bar\varrho$ is the background density at that
epoch).  We can write the linear correlation function of these rare
peaks, at scales where $\frac{\xi(M,r)}{\xi(M,0)} \ll 1$, as
\cite{bbks}  
\begin{equation}
\xi_H(M,r) = \exp\left[ \nu^2 \frac{\xi(M,r)}{\xi(M,0)} \right] - 1
\label{halocorr} 
\end{equation}
Here $\xi(M,r)$ is the correlation function of the smoothed density
field (smoothed at mass scale $M$) evaluated at scale $r$, and $\xi_H$
is the halo correlation function.  The halo correlation has an
exponentially large amplitude over the range of scales where $\nu^2
\frac{\xi(M,r)}{\xi(M,0)} > 1$.  At very large scales, the halo
correlation function is related to the mass correlation as
$\xi_H(M,r) \simeq \nu^2 
\frac{\xi(M,r)}{\xi(M,0)}$.  It follows that at early times,
characterised by $\nu^2/\xi(M,0) > 1$, the halo correlation function
has a much larger amplitude than the mass correlation function at all
scales. 

Time evolution of the linear correlation function for halos, as
described in eqn.(\ref{halocorr}), is controlled by the function
$\nu$~--~the ratio $\frac{\xi(M,r)}{\xi(M,0)}$ being a function of
scale only. 
\begin{equation}
\nu(M,z) = \frac{\delta_c}{\sigma(M,z)} = \frac{D_+(z_\ast)}{D_+(z)} =
\frac{1+z}{1+z_\ast}.
\end{equation}
Here $D_+$\/ is the growing mode for density perturbations in linear
theory and $z_\ast$ is fixed by requiring $\nu(M,z_\ast)=1$.  The last
equality is valid only in the Einstein-deSitter Universe as in that
case $D_+(z) \propto (1+z)^{-1}$.  $D_+$ is a 
monotonically increasing function of time, implying that $\nu$ is a
monotonically decreasing function.  {\it Therefore, at early times,
  the amplitude of correlation function of halos is a rapidly
  decreasing function of time.}  In terms of correlation
bias,\footnote{Note that this bias is different from $\nu$ defined
  earlier.  This degeneracy in nomenclature is unfortunate and we will
  use the term correlation bias for this one to avoid confusion.}
defined as 
\begin{equation}
b^2(r,z) = \frac{\xi_H(r,z)}{\xi(r,z)},
\label{corrbias}
\end{equation}
we can say that at early times, bias $b$ increases more rapidly than
$(1+z)$ with redshift.

Eqn.(\ref{halocorr}) gives only the
linearly extrapolated correlation function for halos.  However, the
qualitative behaviour, being exponentially strong, should survive the
non-linear evolution. 

At later epochs, when $\nu(M) \approx 1$, eqn.(\ref{halocorr}) is no
longer valid.  By this time, most halos of mass $M$ have collapsed and
the halo distribution begins to trace the underlying mass
distribution.  Therefore, further evolution of the halo distribution
must reflect the growth of density perturbations.  There is another way
of explaining this for hierarchical models: low mass halos merge and give
rise to more massive halos.  As gravity brings halos closer for
merger, the halo correlation function must increase.  However, the
rate of growth of correlation function will be slow as {\it
  anti-biased}\/ halos continue to collapse for some time.  In other
words, the correlation bias will change with redshift at a rate slower
than $1+z$.

\begin{figure*}
\hbox{
\epsfxsize=3.3truein\epsfbox[37 26 517 494]{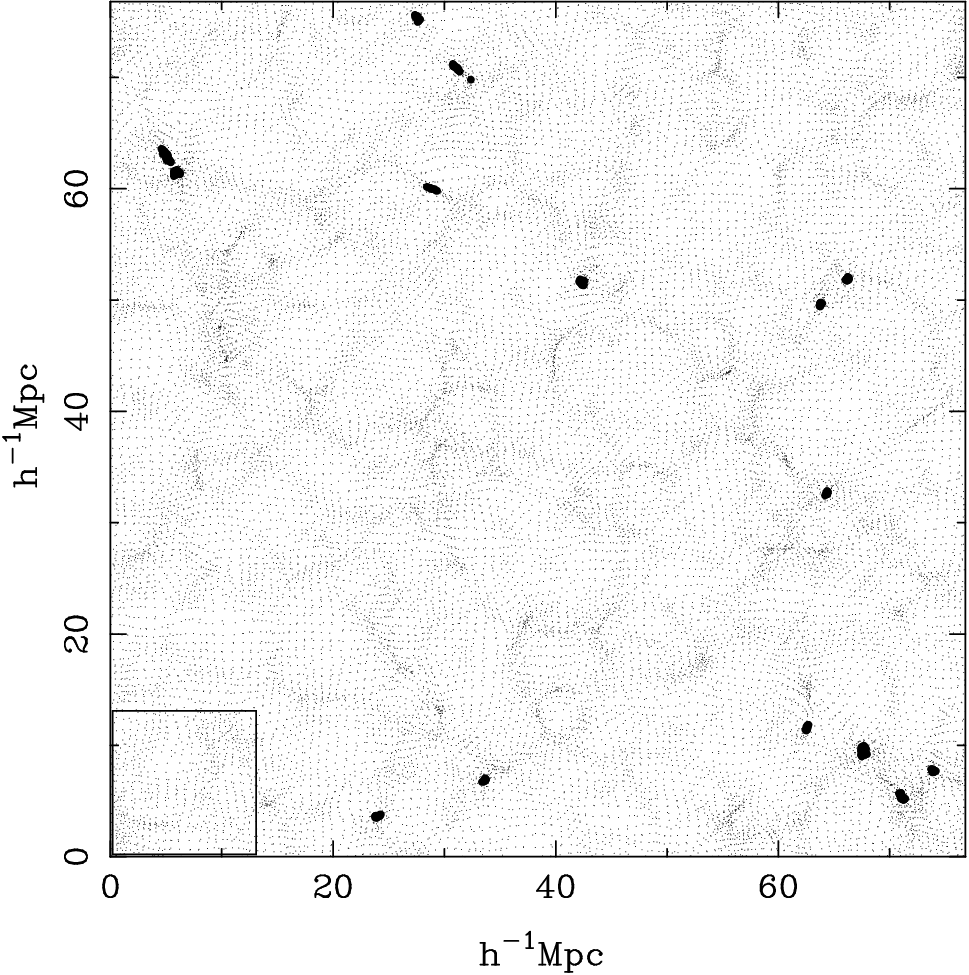}
\epsfxsize=3.3truein\epsfbox[37 26 517 514]{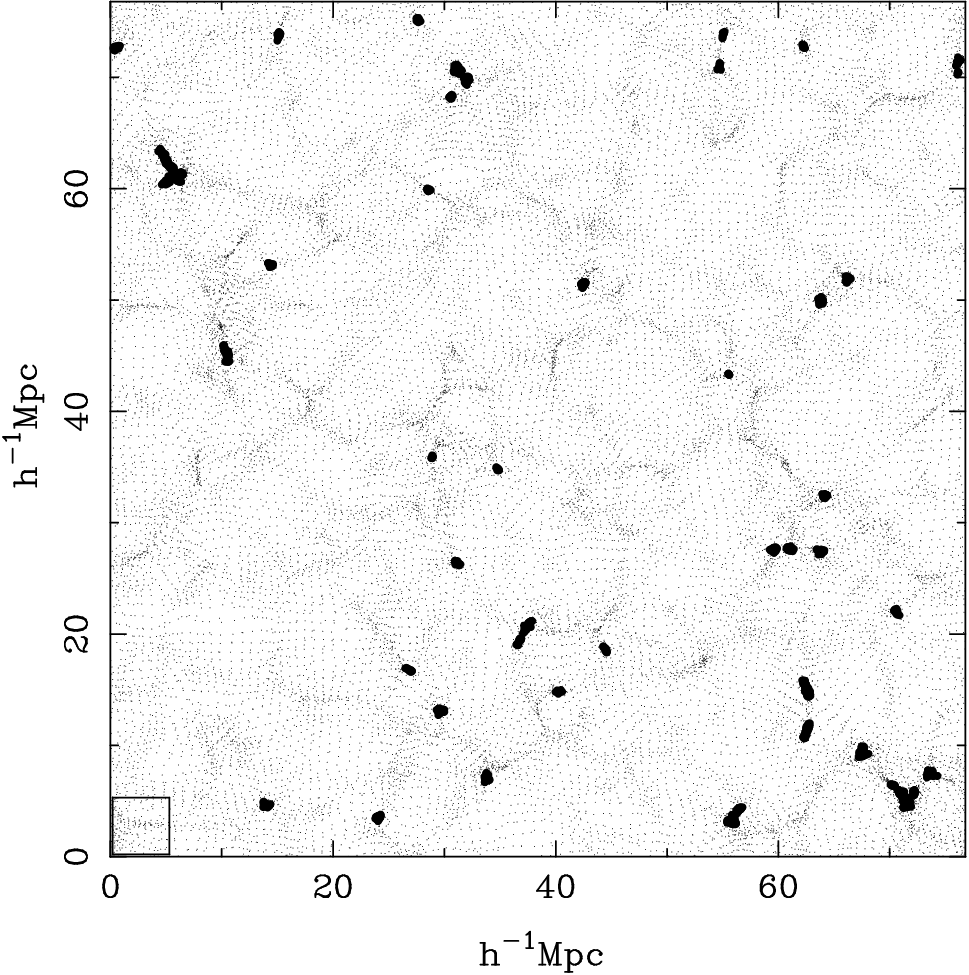}}
\hbox{\epsfxsize=3.3truein\epsfbox[37 26 517 494]{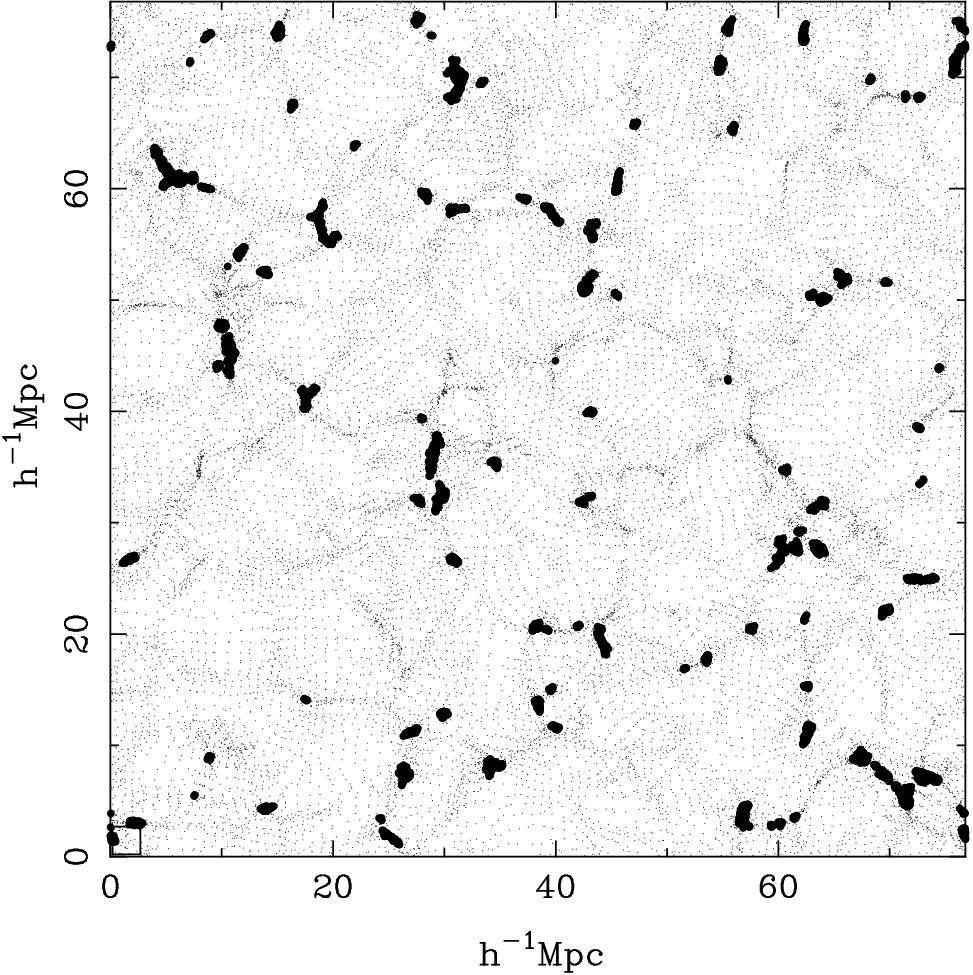}
\epsfxsize=3.3truein\epsfbox[37 26 517 514]{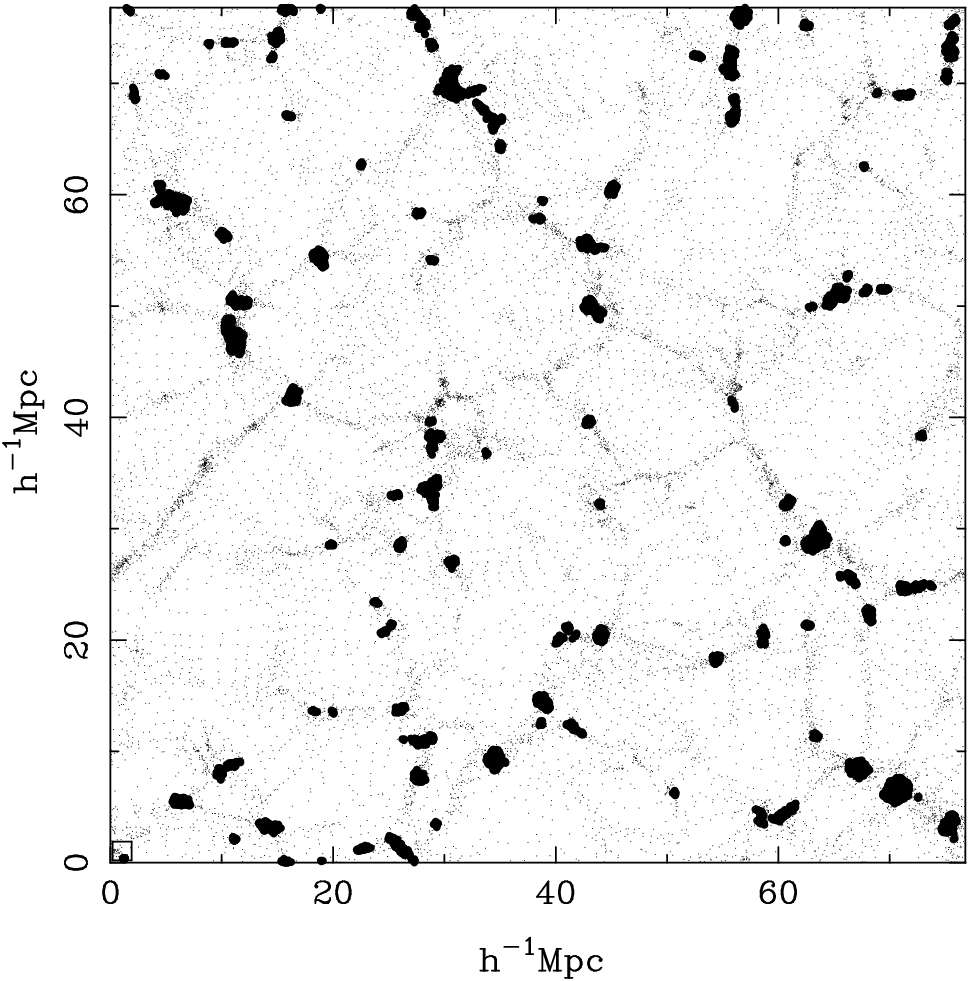}}
\caption{This figure shows the distribution of mass and halos ($M_{halo} \geq
  1.2$~$10^{12}M_\odot$) in a slice from a CDM simulation.  The
  thickness of the slice is $3$h$^{-1}$Mpc.  The halo particles are
  marked as thick points.  To highlight the highly non-uniform
  distribution of halos, we have drawn a box in the lower left corner
  of each frame that shows the region that should contain one halo if
  these were distributed uniformly.  The upper-left panel shows the
  slice at $z=3$, the upper-right panel shows the same at $z=2$, the
  lower-left column shows the same slice at $z=1$ and the lower-right
  panel is for $z=0$.}
\end{figure*}

Thus, there are two effects that contribute to the evolution of
correlation function of halos: The {\it intrinsic}\/ correlation
function of halos that have collapsed, and, gravitational instability.
These effects act in opposite directions.  At early times the rapidly
changing {\it intrinsic}\/ correlation function of the few halos that
have collapsed dominates.  At late times, gravitational clustering
takes over, leading to growth of the correlation function.  
At some intermediate epoch, these effects must cancel each other
leading to a stationary amplitude of the halo correlation function.
This is likely to happen when $\nu \approx 1$.

\section{Numerical Simulations}

We used a set of simulations of the standard CDM model, normalised so
that the linearly extrapolated {\it rms}\/ fluctuations in density,
smoothed with a spherical top hat window function at the scale of
$8$h$^{-1}$Mpc, is $\sigma_8 = 0.6$.  We used $h=0.5$.  All
simulations were done with a PM~(Particle-Mesh) code and $128^3$
particles in a box with the same number of cells.  

We used the friends-of-friends~(FOF) algorithm with a linking length
of $0.3$ grid length to identify dense halos.  This is larger than the
``traditional'' value of $0.2$ and has been chosen so that the mass
function matches the Press-Schechter \cite{psmassfunc} mass function
with $\delta_c=1.68$.  We need to do this to compensate for the lower
resolution of a PM code.  For the purpose of computing the correlation
function, we used halos with $10$ or more particles.  This was done to
avoid noise due to erroneous detections of smaller groups.

In principle, we should identify halos with mass in a given range and
compute the correlation function for these at different epochs.
However, the following reasons force us to use a different strategy:
\begin{itemize}
\item  The FOF algorithm is known to link dynamically distinct halos
  \cite{cdmhalos}.  (Smaller linking lengths tend to dissolve some
  halos.)  This leads to an incorrect estimate of the mass of halos.
  We will also underestimate the number of close pairs of halos, and
  therefore, the amplitude of the correlation function.
\item  Finite mass resolution in numerical simulations leads to the
  over-merging problem \cite{overmerging}.  The over-merging problem
  in collisionless simulations makes it difficult to estimate the
  number of halos of a given mass that may have survived inside bigger
  halos.
\end{itemize}
These problems are particularly serious at late times when the number
of halos is large and there are many cluster sized halos present in
the system.  To circumvent these problems we adopt a different
approach.  We do not consider halos as being one unit each.  Instead,
we compute the correlation function of particles contained in these
halos.  This strategy is also 
appropriate for studying the evolution of galaxy correlation function,
as galaxies are known to survive inside groups and clusters of galaxies.
Although our method ignores the large range in masses of galaxies by
using this particular method, we are able to include the contribution
of very massive halos that are likely to contain many galaxies.

\begin{figure*}
\hbox{
\epsfxsize=3.3truein\epsfbox[41 26 517 494]{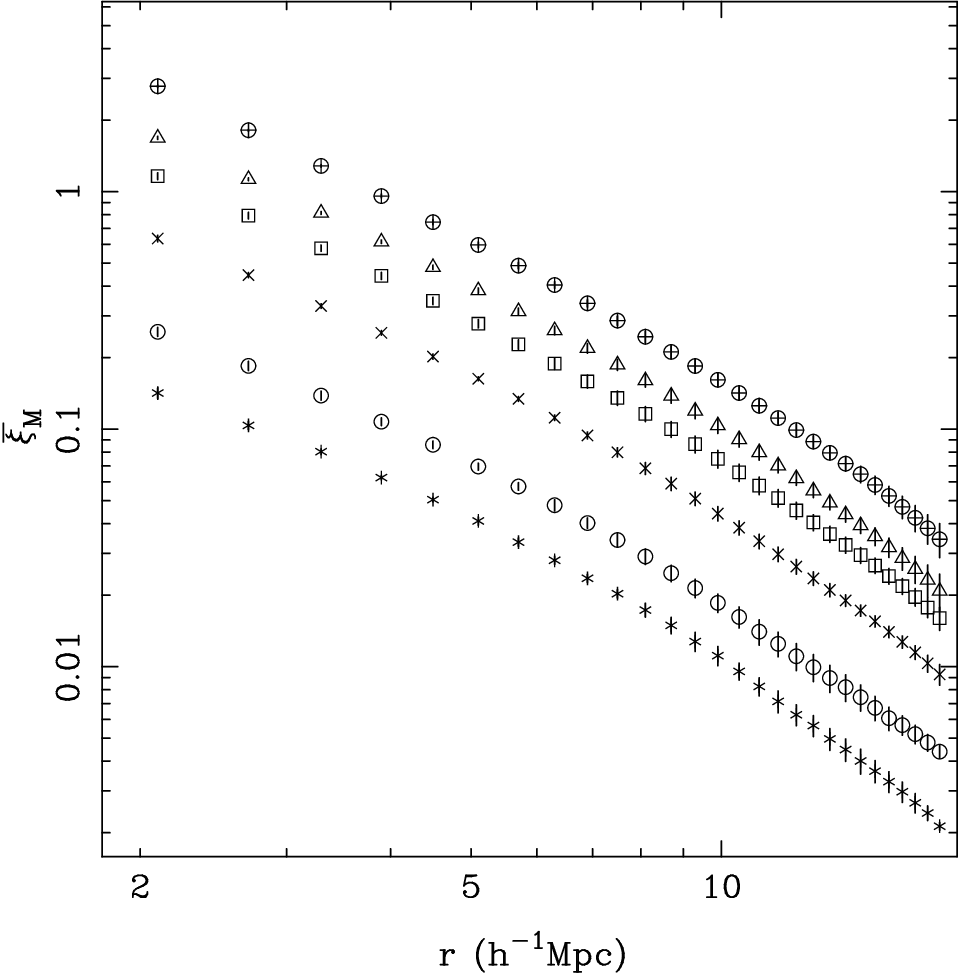}
\epsfxsize=3.3truein\epsfbox[41 26 517 508]{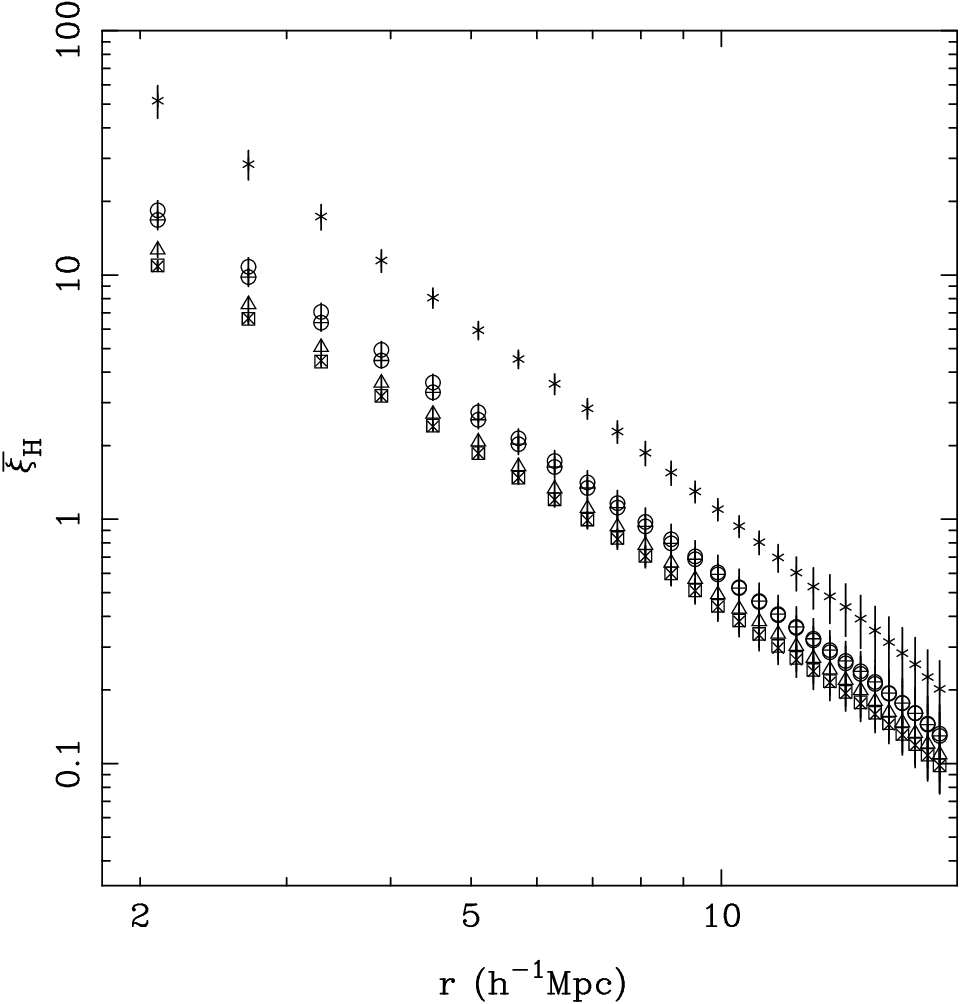}}
\caption{This figure shows the results from simulations of CDM
  spectrum, like the one shown in fig.1.  The averaged correlation
  function $\bar\xi$ is shown as a function of scale
  $r$\/(h$^{-1}$Mpc).  The left panel shows the evolution of clustering
  in the total mass distribution.  As redshift decreases, $\bar\xi$
  increases monotonically.  We have shown the points for $z=3$
  ($\ast$), $z=2$ ($\bigcirc$), $z=1$ ($\times$), $z=0.5$ ($\Box$),
  $z=0.25$ ($\triangle$) and $z=0$ ($\oplus$). The right panel shows
  the correlation function of halos with $M_{halo} \ge 1.2$~$
  10^{12}$M$_\odot$.  The amplitude of the halo correlation function
  decreases rapidly at early times, reaches a minima around $z=1$ and
  then increases again.  The error bars show the $\pm$~$1\sigma$ error
  computed from three realisations of the same power spectrum.}  
\end{figure*}

\section{Results}

To show that halos cluster very strongly at early times, we have
plotted the halos, along with a random subset of all the N-Body
particles, in a slice from a CDM simulation with a box size of
$76.8$h$^{-1}$Mpc.  We have shown the same slice at four redshifts;
$z=3$, $2$, $1$ and $0$.  The thickness of the slice is $3$h$^{-1}$Mpc
(comoving).  The particles in halos ($M_{halo} \geq 1.2$~$
10^{12}$M$_\odot$) are shown as thick dots and the N-Body particles
are shown as thin dots.  A box on the bottom-left corner shows {\it
  the projected area per halo}, if the halos were distributed
uniformly then there should be one halo in a 
region of this size on an average.  It is clear from this figure that
the halo distribution is very non-uniform at early times even though
the underlying mass distribution is fairly homogeneous.  It is also
obvious that the halo distribution develops along the ``skeleton'' of
the late time mass distribution.  Halos form along filaments and
pancakes, or their intersections.  Other simulations also show a
similar pattern of evolution.  (To see the variation with linking
length used for locating halos, see the plots in Brainerd and
Villumsen (1994).)

We describe the clustering properties using the averaged two point
correlation function $\bar\xi(r)$
\begin{equation}
\bar\xi(r) = \frac{3}{r^3} \int\limits_0^r x^2 \xi(x) dx = \frac{3
  J_3(r)}{r^3} 
\label{xibar}
\end{equation}
where $\xi(x)$ is the two point correlation function.  The averaged
correlation function is a measure of the excess neighbours around a
typical point in a sphere of radius $r$, whereas the two point
correlation function is a measure of excess neighbours in a spherical
shell at $r$.

We have plotted the averaged correlation function for the mass
distribution and the halo distribution in fig.2 for simulations like
those shown in fig.1.  The error bars denote the $2\sigma$ dispersion
around the mean evaluated using three realisations of the CDM power
spectrum.  The left panel shows the correlation function for mass and
the right panel shows the correlation function for mass contained
within halos ($M_{halo} \ge 1.2$~$10^{12}$M$_\odot$).  These 
figures show that the clustering in mass increases monotonically, as
expected.  These figures also confirm that the amplitude of the halo
correlation function is large at early times and it decreases very
rapidly.  The amplitude of the halo correlation function reaches a minima
at $\nu \approx 1.0$ and then increases again.  To show this point
more clearly, we have plotted the averaged halo correlation function
at two scales as a function of redshift in fig.3.  Filled squares show
the averaged correlation function at $15$\/h$^{-1}$Mpc and stars show
the averaged correlation function at $5$\/h$^{-1}$Mpc.  A dashed
line and arrows mark the epoch when $\nu=1$ for these halos.  

The fact that bias is high at early times is an ``expected'' result
that is reflected in all the models for evolution of bias (see
below).  However, what we are pointing out here with our simple model,
that is justified by fig.3, is that the bias evolves faster than $1+z$
at epochs where $\nu \gg 1$ and slower than $1+z$ at epochs where $\nu \ll
1$.  This result is contained in the expression for bias derived by Mo
and White (1996) but not in the model by Fry (1996).

In simulations of the CDM model with different box size, we find that
the amplitude of halo correlation function reaches its minimum at
lower $\nu$, i.e. at a later epoch than $\nu =1$, for more negative
indices.  We find the range $\nu_{min} = 0.96$ ($n_{eff} = -2.4$) to
$\nu_{min} = 1.1$ ($n_{eff} = -1.9$).  Here, the index $n_{eff}$ is
defined as
\begin{equation}
n_{eff} = -3 + \frac{\partial \ln \sigma^2\left(r\right)}{\partial\ln
  r} \label{index}
\end{equation}
where the derivative is evaluated at the mass scale of halos.  We are
studying the variation of $\nu_{min}$ with the index for power law
models. 

The shape of both the mass correlation and halo correlation changes
with time.  The halo correlation function becomes less steep with
time, whereas the mass correlation function becomes steeper.  If we
compare the shape and amplitude of the halo and mass correlation
function at $z=0$ in fig.2 then it is apparent that the halo
correlation function has a higher amplitude and it is steeper than the
mass correlation function.  To quantify this, we have plotted the
correlation bias as a function of scale at four redshifts in fig.4.
The correlation bias is defined here using the averaged correlation
function in the manner of eqn.(3).  The left panel shows these for ``mass
weighted'' halo correlation function, i.e. correlation function of
mass in halos.  The right panel shows bias 
computed from unweighted halo-halo correlation function.  These differ
at small scales where the effect of over merging is important.  At late
times, the lower curve shows strong suppression of bias at small
scales due to this effect.  However, at early times, this effect is
relatively unimportant and the unweighted halo correlation function
gives a better estimate of bias.  At late times bias computed from the
correlation function of mass in halos (left panel) should resemble
the galaxy correlation function.  The curve corresponding to $z=0$
shows that bias is a function of scale and it increases at small
scales.  At large scales it approaches an asymptotic value and does
not change much beyond $10$\/h$^{-1}$Mpc.

The qualitative behaviour of the correlation bias as a function of
scale does not evolve very strongly.  However, at early times the
increase in bias at 
smaller scales is more prominent than at late times.  This is a
reflection of the change in shape of the matter and halo correlation
functions.  The scale dependence of correlation bias has
important implications for inversion of galaxy power spectrum to
obtain the initial power spectrum (see \S{5.1}). 

Fig.3 shows the variation of correlation function for halos at two
comoving scales ($5$h$^{-1}$Mpc and $15$h$^{-1}$Mpc) as a function of
redshift.  To compare with theoretical models, we first describe the
figure in terms of correlation bias.  At early times bias varies at a
rate faster than $1+z$, the rate of variation is close to $1+z$ at
around $\nu=1$ and then it slows down at lower redshifts.  In other
words, bias varies very rapidly at early times but at late times the
variation is very slow.  

The first model we consider assumes that all the halos are assumed to
form at the same redshift $z_\ast$.  These halos form with some
initial local bias\footnote{Local bias is defined as the ratio
  between density contrasts in galaxy number and the underlying mass
  distribution.  The model by Fry (1996) assumes that the local bias is
  independent of position. The assumption of local bias implies a
  correlation bias that is independent of scale.} $b_\ast$.  The model
then evolves the clustering of these halos by solving the continuity
equation.  For comparison, we have plotted (solid lines) the
prediction of this model through the low redshift points.  The model
predicts 
\begin{equation}
b = \frac{z_\ast + b_\ast}{1 + z_\ast} + z\frac{b_\ast - 1}{1 +
  z_\ast} = 1 + (1 + z) (b_0 -1 ).
\end{equation}
$b_0$ is the bias the $z=0$.  As is clear from this equation, an
unbiased set of tracers always remain unbiased.  At early times, the
combined assumption of conservation of the numbers of halos and a
local bias is not valid as the rate of formation of halos is high and
these are not fair tracers of the mass distribution~--~this leads to
the strong disagreement of predictions with the simulation data at
early times.  

\begin{figure}
\epsfxsize=3.3truein\epsfbox[42 30 502 503]{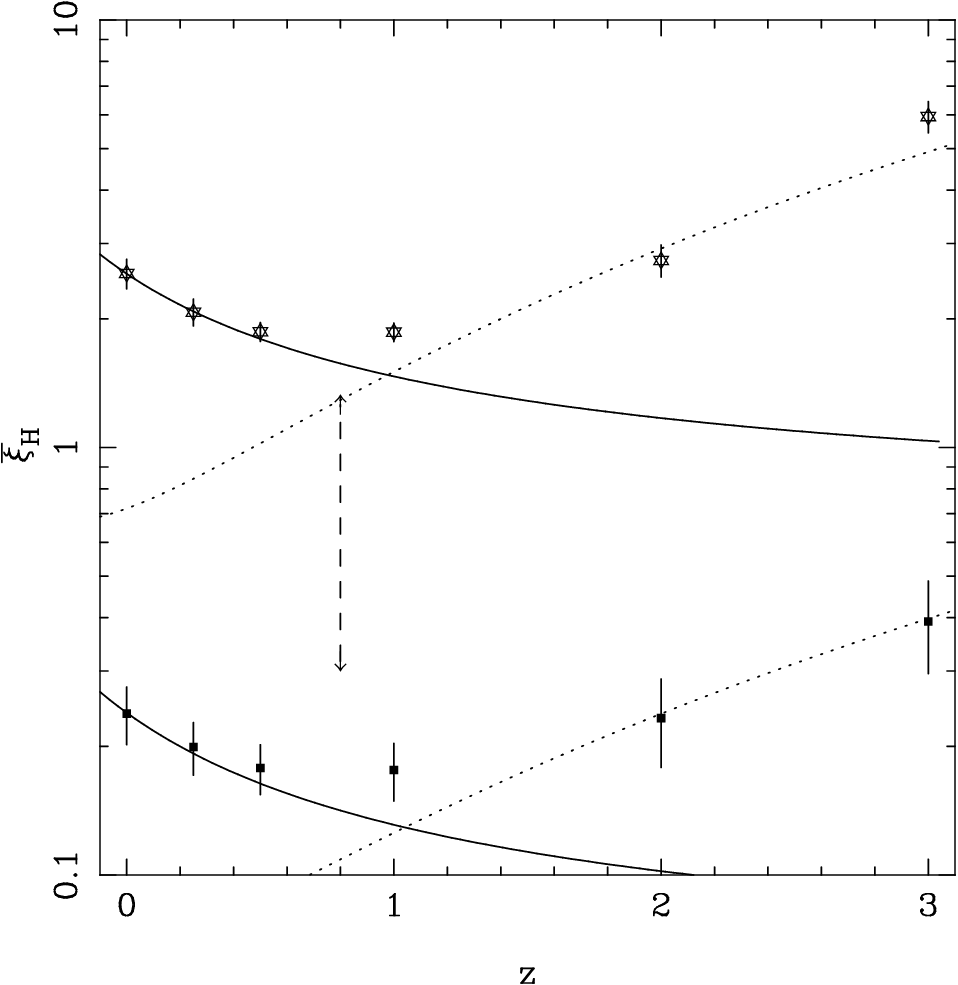}
\caption{This plot shows the averaged halo correlation function at two
  scales as a function of redshift.  Filled squares show the averaged
  correlation function at $15$\/h$^{-1}$Mpc and stars show the averaged
  correlation function at $5$\/h$^{-1}$Mpc.  Error bars depict the
  $\pm 1\sigma$ dispersion around the mean computed from three
  realisations of the power spectrum.  A dashed line and
  arrows mark the epoch when $\nu=1$ for halos of mass $M_{halo} \ge
  1.2$~$10^{12}$M$_\odot$.  It is clear that the minimum amplitude is
  reached near that epoch.  Thick lines mark the predicted evolution
  of halo correlation function in the model by Fry (1996).  Dotted
  lines mark the evolution of bias in the model of Mo and White (1996)
  where we have used the expression for effective bias from Matarrese
  et al. (1997).} 
\end{figure}

Mo and White (1996) have developed a model for biasing in which the
correlation bias is assumed to be independent of scale for halos of
any given mass.  They predict variation of the form
\begin{equation}
b = 1 + \frac{1}{\delta_c} \left( \nu^2 -1 \right) = 1 +
\frac{1}{\delta_c}\left\{ \left(\frac{1+z}{1+z_\ast}\right)^2 - 1
\right\} 
\end{equation}
where the second equality holds only for an Einstein-deSitter Universe.  

\begin{figure*}
\hbox{\epsfxsize=3.3truein\epsfbox[43 26 517 494]{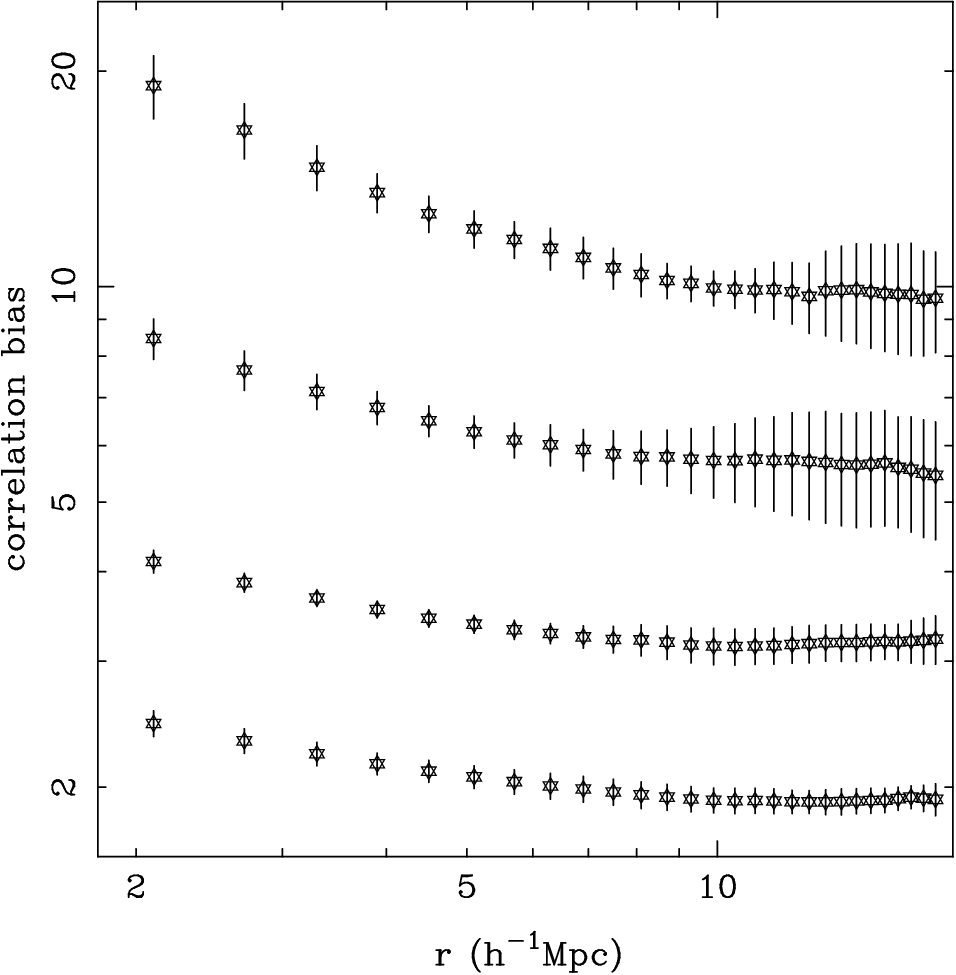}
\epsfxsize=3.3truein\epsfbox[43 26 517 494]{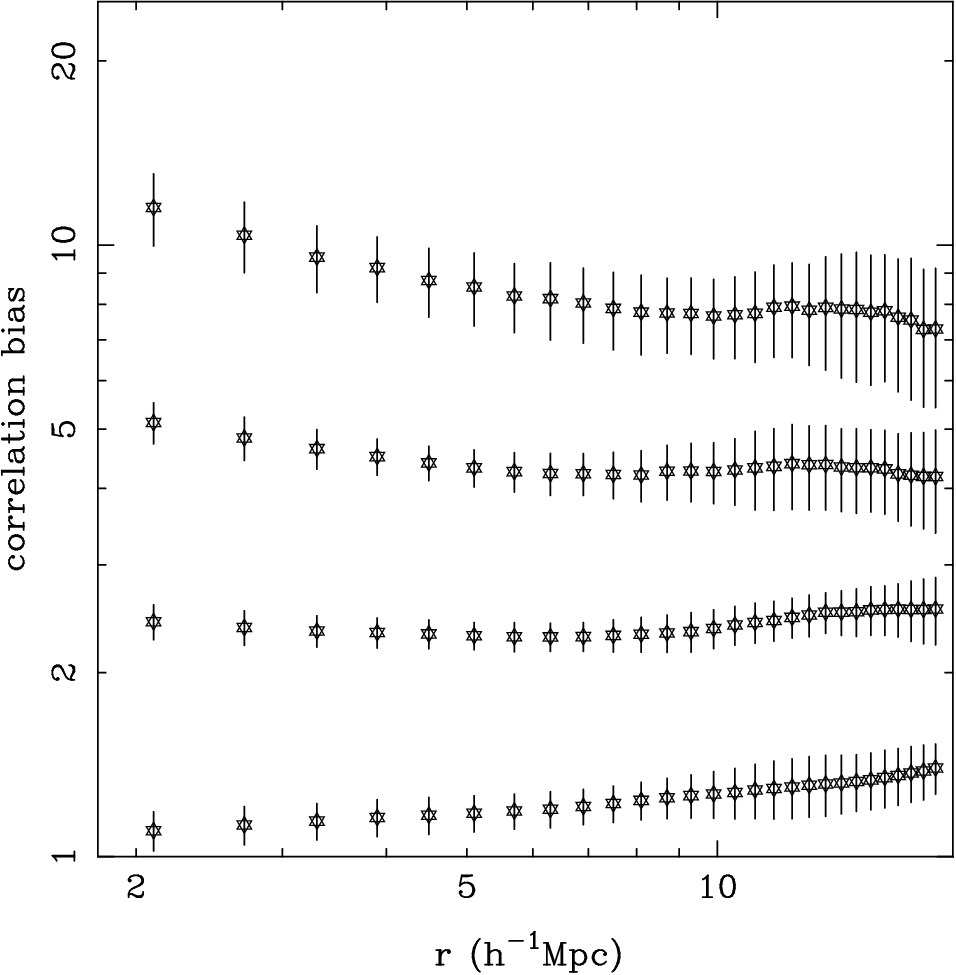}}
\caption{This figure shows correlation bias as a function of scale for
  four redshifts.  This is shown for halos of mass $M_{halo} \ge
  1.2$~$10^{12}$M$_\odot$ at $z=3$, $2$, $1$ and $0$.  Bias decreases
  monotonically with redshift so the points for $z=3$ are the ones at
  the top and the points for $z=0$ are at the bottom.  Error bars
  depict the $\pm 1\sigma$ dispersion around the mean computed from
  three realisations of the power spectrum.  The left panel shows
  these for mass weighted halo correlation function.  The right panel 
  shows bias computed from unweighted halo-halo correlation function.
  These differ at small scales where the effect of over merging is
  important.  At late times, the lower curve shows strong suppression
  of bias at small scales due to this effect.}
\end{figure*}

Matarrese et al. (1997) have generalised the model of Mo and White
(1996) by combining the bias for different mass scales.  They use many
different ansatz for variation of ``visibility'' of halos and compute
the correlation bias in each case.  In general, they
get the following form for correlation bias 
\begin{equation}
b(z) = c_1 + c_2 (1 + z)^\beta
\end{equation}
where some of the parameters can be fixed/related by choosing an
appropriate ansatz.  Using the coefficients for the appropriate
$M_{min}$~$=(M_{halo})$ and accounting for the fact that we are using
a different 
normalisation, we have computed the predicted amplitude of correlation
function for comparison with the values obtained from simulations.
The predictions of their model are shown as dotted lines in fig.3.  It
is clear that though the match is good at high redshifts, this model
does poorly at low redshifts.  The main reason being that the parent
model, Mo and White (1996), deals with the unweighted
halo correlation function.  Therefore any comparison with a weighted
halo correlation function will not work.  However, as we argued in
\S{3}, the mass weighted correlation function is a better candidate
for the galaxy correlation function.  Unweighted halo correlation
function predicts an anti-bias at late times.  This is never seen in
the weighted correlation function and hence the large mismatch.

Some authors have constructed analytical models to understand the
evolution of bias (Catelan et al. 1997 and the references cited therein)
field where the concept is generalised from statistical bias (only a
function of epoch) to a bias that depends both on position and epoch.
In these models the mapping from the initial halo distribution to the
final one is done using perturbative or approximate methods.  

\section{Discussion}

In \S{4} we described evolution of the
correlation function for mass contained in halos of mass greater than
a given cutoff.  These results, when applied to galaxies, have many
important implications.  In \S{5.1}, we will discuss the calculation
of the initial power spectrum from the observed galaxy correlation
function in view of the results presented in \S{4}.  In \S{5.2} we
turn to the question of evolution of galaxy/quasar clustering and its
relation with the evolution of halo clustering.  Lastly, we outline
some implications of these results for evolution of the inter-galactic
medium and galaxy formation models in \S{5.3}.

\subsection{Galaxy Correlation Function and the Initial Power Spectrum}

In this section, we assume that the halo distribution and galaxy
distribution are the same at the present epoch.  This is a reasonable
assumption for studying galaxy clustering at a given epoch, as long as
the mass of halos is not too different from the mass of typical
galaxies studied in surveys. 

The shapes of mass and galaxy correlation functions are different,
even at late times (fig.2, fig.3 and fig.4).  These differences
introduce errors in calculation of the initial power spectrum from
observations of galaxy clustering using scaling relations
\cite{pd96}. 

Fig.5 shows the non-linear index $n_{nl}$ as a function of the linear
index $n_{lin}$ of the averaged correlation function.  We define
$n_{nl}$ as in eqn.(\ref{index}) except that $\sigma^2$ is replaced by
$\bar\xi$.  This relation between the indices is obtained by using the
power law fit \cite{crit} in the quasi-linear regime ($1 \le \bar\xi
\le 200$)\footnote{For a power law correlation function, $\bar\xi = 3
  \xi /(3-\gamma)$.  Thus for $\gamma = 1.8$, as suggested by
  observations, this relation between the indices can be used up to
  about $8$h$^{-1}$Mpc.} to the scaling relation between the linear
and the non-linear correlation function \cite{hamil}.  This figure
shows that this relation flattens out for indices above $n_{nl}=-1$.
Two reasons contribute to this flattening: 
\begin{itemize}
\item The index $n_{lin}=-1$ is a ``critical index'' in the sense that
  power spectra with an index close to $n_{lin}=-1$ change shape so that
  $n_{nl}$ is closer to this critical index in the quasi-linear
  regime \cite{crit}. 
\item Gravitational clustering of hierarchical models ($n > -3$)
  restricts the non-linear indices below $ n_{nl} \le 3/\bar\xi$
  \cite{crit}.  In other words, the non-linear index of the power
  spectrum must be in the range $-3 \ge n_{nl} \ge 0$, irrespective of
  the initial index.
\end{itemize}
Gravitational instability acts to decrease (but not to erase)
the differences between different initial conditions.  Therefore, any
uncertainties in the non-linear mass correlation function translate
into much larger uncertainties in the initial power spectrum.  The
difference in the shape/slope and the amplitude of the galaxy
correlation function and the mass correlation function is one such
uncertainty. 

\begin{figure}
\epsfxsize=3.3truein\epsfbox[46 114 505 404]{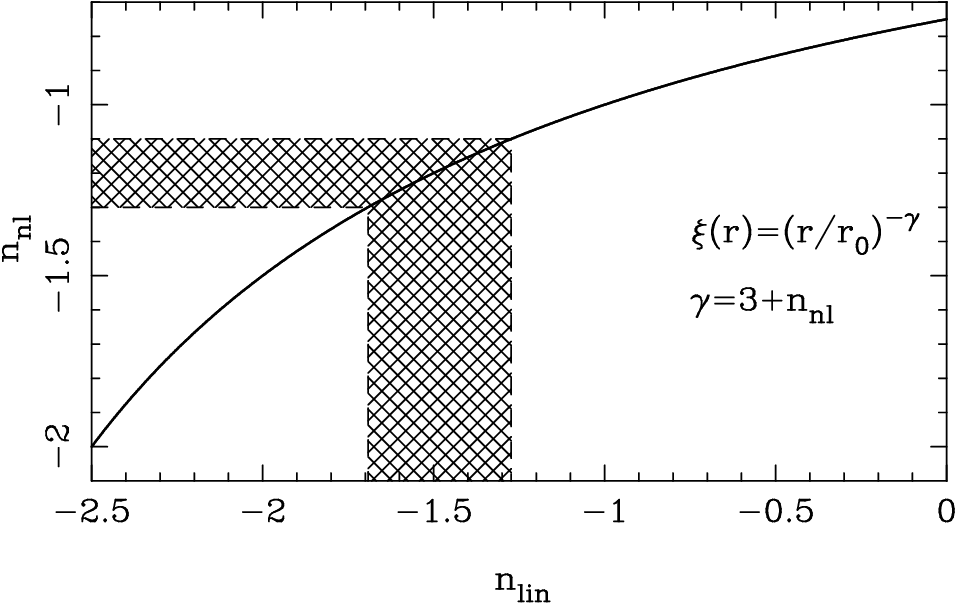}
\caption{This figure shows the non-linear index of the power spectrum
  as a function of the linear index in the quasi-linear regime ($1 \le
  \bar\xi \le 200$).  This relation is derived using a power law fit
  to the relation between the linear and evolved correlation (Hamilton
  et al. 1991).  We have shown a horizontal shaded region to mark the
  region $\gamma = 1.8 \pm 0.1$.  This translates into a linear index
  of $\gamma_{lin}\simeq 1.5 \pm 0.2$, marked by the vertical shaded
  region.  This demonstrates the amplification of uncertainty in
  the index of power spectrum in the inversion process.}
\end{figure}

In order to assess this amplification of uncertainty in a quantitative
manner, we have mapped a narrow range of non-linear indices,
$n_{nl}=-1.2 \pm 0.1$ ($\gamma = 1.8 \pm 0.1$), to the corresponding
linear indices.  The 
permitted range of linear indices is much larger ($n_{lin} \simeq -1.5
\pm 0.2$; $\gamma \simeq 1.5 \pm 0.2$).  Therefore, even a small error
in the non-linear index\footnote{If we assume the difference between
  halo clustering and mass clustering to be the only source of error,
  then we can estimate its magnitude from simulations.  The curves for
  $z=0$ in fig.2 show this difference to be $\Delta n \simeq 0.2$ for
  $M_{halo} \ge 1.2$~$10^{12}$M$_\odot$, twice the uncertainty of
  $0.1$ we used in our example.} of the power spectrum can lead to a
much larger error in the linear power spectrum.  In addition to the
uncertainty in slope, uncertainty in amplitude also contributes to the
error.

The above discussion shows that inverting the galaxy correlation
function to determine the initial power spectrum leads to
amplification of uncertainties.  This is especially true of scales in
the non-linear regime where, in general, correlation bias depends on
scale.  At larger scales, bias is expected to be scale independent and
the main source of error will be the uncertain difference in amplitude. 

\subsection{Halos and Galaxies}

Discussion in this paper has assumed, so far, that galaxy clustering
and halo clustering are the same thing.  Although this is a reasonable
assumption if we are interested in only one epoch with an appropriate
choice of minimum halo mass, the same is not true for the relative
evolution of the two distributions.  Here, we list some plausible
alternatives and outline the evolution of galaxy correlation function
for each one. 

At any given epoch, there are more low mass galaxies than high mass
galaxies.  This suggests that the galaxy correlation function is
dominated by the smallest galaxies\footnote{There is, of course, the
  problem of identifying the {\it same}\/ population of galaxies at
  different redshifts}.  In such a case, the galaxy correlation
function must evolve in a manner similar to the halo correlation
function and its amplitude must go through a minimum at some epoch.
For low mass galaxies, this minimum could be at $z \gg 0$.  

The assumption outlined above applies in a much better way to quasars
as these cannot form in very low mass halos and the redshift at which
the amplitude of quasar correlation reaches the minima may be low
enough to be observable.  If the minimum halo mass 
associated with quasars is greater than $10^{11} M_\odot$ then the
observed quasar correlation function should have a higher amplitude at
$z >2$ than at $z < 1$.  Considering the small probability of a given
halo hosting a quasar at a given time, the more appropriate measure
for clustering of quasars is provided by the unweighted halo
correlation function.

Recent estimates of the quasar correlation
function, though based on small datasets, show that clustering of
quasars is indeed stronger at higher redshifts (Kundic 1997; La
Franca, Andreani and Cristiani 1997).  Authors of the second paper
claim that the observed evolution of quasar clustering cannot be
explained in all models of AGN activity.  Our analysis suggests that
this need not be true.  Quasars must show decreasing clustering
amplitude if the AGN activity is not correlated with the large scale
environment.

The assumption of a mass threshold also applies very well to
groups and clusters of galaxies.  Therefore, the cluster correlation
function should show a higher amplitude at higher redshifts.

We can use a different ansatz: we can assume that the brightest
galaxies reside in halos that have formed relatively recently.  We can
identify the mass scale $M_\ast(z)$ by assuming that these correspond to
$\nu(M_\ast) \simeq 1$.  For hierarchical models, we know that
$\sigma(M_1,z) > \sigma(M_2 > M_1,z)$, implying that $\nu(M_2,z) >
\nu,M_1)$.  As a high $\nu$ corresponds to a high correlation bias, it
is clear that the halo correlation function with threshold $M_1$ will
have a lower amplitude than the halo correlation function with
threshold $M_2$.  This indicates that the correlation function of
$M_\ast$ halos at $z_1$ will have a much lower amplitude than its
equivalent at $z_2$.  Therefore we expect rapid evolution of the
galaxy correlation function, much faster than the linear growth rate,
if galaxies correspond to $M_\ast$ halos.  This may be a good model
for evolution of clustering of faint galaxies \cite{evclus}. 

The above discussion shows that the evolution of galaxy clustering is
likely to depend strongly on our choice of relation between halos and
galaxies at different redshifts.  If we identify similar objects at
different redshifts, then these will show either a decreasing
correlation function or nearly constant clustering in comoving
co-ordinates.  On the other hand, if we are working with the most
prominent/numerous objects at each redshift then we should see rapid
evolution of clustering.  Gastrophysical effects, in general, make the
relation of halos and galaxies a little fuzzy, and this will result in
an uncertain linear combination of the two types of evolution of
clustering.   

Considering the uncertain relation of the relative evolution of
galaxy/halo and mass clustering, the rate of growth of the correlation
function at large scales should not be interpreted as the linear
growth rate of density perturbations.  Direct determination of
cosmological parameters by assuming the two rates to be identical
can lead to wrong results.

\subsection{Galaxy Clustering and Reionisation of the IGM}

Observations show that the Inter-Galactic Medium (IGM) is ionised at
the highest redshifts we can probe using known quasars and galaxies.
Scenarios for reionisation of the IGM fall in two basic
categories:~Ionisation by quasars, which have a low number density and
hence the IGM has a very patchy structure at early times.  Ionisation
by proto-galaxies, dwarf galaxies or star clusters that are
distributed uniformly, leading to a quick and homogeneous
reionisation.  

We have shown in this paper that halos, irrespective of their mass,
cluster very strongly at early times.  Therefore, the reionisation of
the IGM will be patchy at a scale much larger than $n^{-1/3}$, where
$n$ is the number density of the sources of ionisation.  The
non-uniformity of halos at early times is illustrated in fig.1.  

It is thought that the earliest clusters of stars formed when low mass
halos $M_{halo} \simeq 10^6 M_\odot$ collapse after cooling by H$_2$
line cooling \cite{molcool}.  If these clusters of stars were
responsible for reionisation of IGM, then, as these have a very large
number density, $n^{-1/3} \approx 10$kpc$_{proper}$, the ionisation
structure of the IGM will be fairly homogeneous.  However, UV
radiation from the first clusters leads to dissociation of H$_2$
molecules and hence the Jeans mass increases by a considerable amount
\cite{destrmol}.  The second generation of collapsed object are like
dwarf galaxies $M_{halo} \simeq 10^{8-9} M_\odot$ and these have a
slightly lower number density, $n^{-1/3} \approx 0.1$Mpc$_{proper}$.
A visual comparison with fig.1 suggests that some parts of the
Universe will be at a much greater distance from the nearest source of
ionising radiation.  Therefore, one may expect patchiness at scales of
$1$Mpc$_{proper}$, or about $10$Mpc$_{comov}$.   The early stages of
reionisation from such sources may be observed using the $21$cm
tomography type of observations, though at a scale smaller than that
suggested for quasars \cite{tomo}.  On the other hand, if quasars are
responsible for ionising the IGM then the scale of patchiness will be
much bigger than that expected from the number density.

The relative distribution of halos and the underlying mass in fig.1
suggests that halos form preferentially along filaments/pancakes or
their intersections.  Denser filaments and pancakes can ``resist''
being ionised much more efficiently than the under dense regions.
Photo-ionisation in the under-dense region will increase the Jeans
mass and suppress collapse of low mass halos \cite{supress} at late
times.  This may be the reason for the apparent lack of dwarf
galaxies in voids.  This effect will not be as prominent in the
filaments as the recombination time is smaller, and, ionisation
proceeds much more slowly along over-dense regions than it does in
the under-dense regions.

\section{Conclusions}

In \S{5}, we have highlighted some implications of the model and
simulation results presented in this paper.  Our conclusions are
summarised below:
\begin{itemize}
\item  The halo correlation function for halos with mass greater than
  ($M_{halo} =1.2$~$10^{12}M_\odot$) at $z=3$ is higher than the
  correlation function of halos above the same threshold today.
  Therefore, it is only to be expected that the galaxy correlation
  function at high redshifts should have a large amplitude
  \cite{wallobs,walls}. 
\item  The halo correlation function (and the correlation function of
  mass contained in halos) has a different shape, as compared to the
  non-linear correlation function of mass.  Therefore, inverting the
  galaxy correlation function to obtain the linear correlation
  function, or the power spectrum, for mass can lead to wrong answers
  at small scales ($l \le 8$h$^{-1}$Mpc).
\item  The growth rate of correlation function should not be used,
  directly, to compute omega/lambda by assuming linear growth or
  stable clustering - we have shown that the rate of evolution of halo
  correlation function has little in common with the matter
  correlation function, and does not depend in any obvious way on
  cosmological parameters.  
\item Quasar correlation function, as it corresponds to high mass
  halos, should show a decreasing phase at $z>1$.
\item Cluster correlation function should also show a decreasing
  phase, if we can define a sample of clusters at higher redshifts
  ($z\simeq 0.5$) with the same mass threshold.
\item The rapid evolution of clustering of faint galaxies can be
  understood if these correspond to $M_\ast$ galaxies at high
  redshifts.
\item Earliest structures to form in the Universe will cluster
  strongly, therefore the sources that reionise the Universe will
  occur in groups that are further apart than expected from their
  number density and a nearly uniform distribution.  This implies that
  the reionisation will be very patchy and it may be possible to
  observe the patchiness using the redshifted $21$cm radiation
  \cite{tomo}.
\end{itemize}

\section*{ACKNOWLEDGEMENT}

I would like to thank Martin Rees, K.Subramanian, Ofer Lahav and Shiv
Sethi for many useful discussions.  I acknowledge the support of PPARC
fellowship at the Institute of Astronomy.

\label{lastpage}

\end{document}